\documentclass[%
 reprint,
 amsmath,amssymb,
 aps,
]{revtex4-1}

\usepackage{graphicx}% Include figure files
\usepackage{dcolumn}% Align table columns on decimal point
\usepackage{bm}% bold math
\begin{document}

\preprint{APS/123-QED}

\title{Broadening near-field emission for performance enhancement in thermophotovoltaics}% Force line breaks with \\

\author{Georgia T. Papadakis}
\affiliation{Department of Electrical Engineering, Ginzton Laboratory, Stanford University, California, 94305, USA}
\author{Siddharth Buddhiraju}
\affiliation{Department of Electrical Engineering, Ginzton Laboratory, Stanford University, California, 94305, USA}
\author{Zhexin Zhao}
\affiliation{Department of Electrical Engineering, Ginzton Laboratory, Stanford University, California, 94305, USA}
\author{Bo Zhao}
\affiliation{Department of Electrical Engineering, Ginzton Laboratory, Stanford University, California, 94305, USA}
\author{Shanhui Fan}\email{shanhui@stanford.edu}
\affiliation{Department of Electrical Engineering, Ginzton Laboratory, Stanford University, California, 94305, USA}

\begin{abstract}
The conventional notion for achieving high efficiency in thermophotovoltaics (TPVs) is to use a monochromatic emission at a photon energy corresponding to the band gap of the cell. Here, we prove theoretically that such a notion is only accurate under idealized conditions, and further show that when non-radiative recombination is taken into account, efficiency improvement can be achieved by broadening the emission spectrum, due to an enhancement in the open-circuit voltage. Broadening the emission spectrum also improves the electrical power density, by increasing the short-circuit current. To practically illustrate these findings, we focus on surface polariton-mediated near-field TPVs. We propose a versatile design strategy for broadening the emission spectrum via stacking of multiple plasmonic thin film layers. As an example, we consider a realistic ITO/InAs TPV, and predict a conversion efficiency of $50\%$ simultaneously with a power density of nearly $80$ W$/\mathrm{cm}^2$ at a $1300$ K emitter temperature. The performance of our proposed system far exceeds previous works in similar systems using a single plasmonic layer emitter.
\end{abstract}

\maketitle

%\tableofcontents

\par{A thermophotovoltaic (TPV) system is a solid-state renewable energy approach that is of potential importance for a number of applications including solar energy harvesting, and waste heat recovery. In these systems, a photovoltaic (PV) cell is separated from a thermal emitter by a vacuum gap. The PV cell converts the thermal radiation from the emitter into electricity. Since far-field thermal radiation is fundamentally limited in its power density by the blackbody limit, there are significant theoretical \cite{Chen_NFTPV2003,Greffet_NFTPV2006,Park_NFTPV2008,Ilic_TPV2012, Lipson_NFTPV2016, Zhao_thermophotonic2018,Abdallah_ReviewNFTPV} and experimental \cite{Reddy_NFTPV2018} efforts in exploring near-field TPV systems. In these systems, by reducing the spacing between the PV cell and the thermal emitter to be smaller than the relevant thermal wavelength, the power density can far exceed that in the far-field system.}

\par{In standard analysis for TPV systems, assuming an idealized PV cell without non-radiative recombination, it is known that the efficiency of the TPV maximizes to the Carnot efficiency limit when the thermal exchange spectrum between the emitter and the PV cell has a narrow-band, located at the band gap of the PV cell \cite{Greffet_NFTPV2006,Ideal_NFTPV_Jabob}. Motivated by this analysis, significant efforts have been devoted to develop near-field TPV systems where the emitters support surface plasmon \cite{Greffet_NFTPV2006, Ilic_TPV2012, Palasantzas_NFTPV2014, Lipson_NFTPV2016, Joannopoulos_SqueezingNFHT,Francoeur_NFTPV2017, Zhao_NFTPV} or phonon \cite{Chen_NFTPV2003} polaritons, and hence the thermal exchange spectrum between the emitter and the PV cell is narrow-banded \cite{NFTPV_review2008,Abdallah_ReviewNFTPV,Greffet_NFTPV2006}.}

\par{In this paper, we provide a theoretical analysis of a TPV system, where the PV cell has significant non-radiative recombination, as is typical for most PV cells assumed in previous analysis of near-field TPV systems. We show that in the presence of significant non-radiative recombination, starting from the narrow-band thermal emission limit as discussed above, the efficiency in fact \textit{increases} as the bandwidth of the thermal exchange increases. Since in general the power density of the TPV system should increase as the bandwidth the thermal exchange increases, our results indicate that increasing the bandwidth of the thermal exchange in fact increases \textit{both} the efficiency and the power density of TPV systems. Motivated by this theoretical analysis, we introduce a design of a broad-band near-field thermal emitter, where we introduce a spatial gradient of the plasma frequency. We show that such a design of a broad-band near-field thermal emitter significantly increases the power density and efficiency of a TPV system, as compared with the use of a standard surface plasmon near-field emitter.} 

\par{We start by considering a detailed balance analysis of a TPV system with an emitter at temperature $T_\mathrm{H}$ and a cell at temperature $T_\mathrm{C}$ \cite{DetailedBalance1961}. The current density $J(V)$ of the PV cell is:
\begin{equation}\label{eq:1}
J(V)=J_\mathrm{e}-J_\mathrm{o} e^{qV/kT_\mathrm{C}}+R_\mathrm{o}-R(V)
\end{equation}
where $V$ is the operating voltage of the cell. The first and second terms correspond, respectively, to the radiative generation and recombination of electron-hole pairs, whereas the third and fourth terms correspond to non-radiative generation and recombination, respectively. $J_\mathrm{o}$ is the current arising from the thermal emission of the cell, whereas $J_\mathrm{e}$ arises from the absorption of the cell in the frequency range above the band gap. These are given by:
\begin{equation}\label{eq:2}
J_\mathrm{o/e}=\frac{q}{4\pi^2}\int_{\omega_\mathrm{g}}^{\infty} \Phi(\omega) n(\omega,T_\mathrm{C/H}) d\omega
\end{equation}
where $n(\omega,T)=(e^{\hbar\omega/kT}-1)^{-1}$ is the Planck distribution for photons with energy $\hbar\omega$ at temperature $T$, $\omega_\mathrm{g}$ is the band gap, and $\Phi(\omega)$ is the normalized emission spectrum. In Eq. \ref{eq:2} we considered the case where radiative exchange occurs exclusively between the cell and the emitter, i.e. there is no radiation leackage to the environment. Based on the standard treatment with fluctuational electrodynamics \cite{Rytov_Book,PolderNFHT,CarminatiGreffet_PRL2000, Raschke_NFHTReview, Basu_NFHTtreatment}, $\Phi(\omega)$ is given by: 
\begin{equation}\label{eq:3}
\Phi(\omega)=\int_{0}^{\infty}\xi(\omega,\beta)\beta d\beta
\end{equation}
where $\xi(\omega,\beta)$ is the probability, summed over the two polarizations, for a photon with frequency $\omega$, and in-plane wavenumber $\beta$, to be transmitted through the vacuum gap.}

\par{The efficiency of a TPV system is defined as the ratio $\eta=(P_\mathrm{el}/P_\mathrm{phot})\times100\%$ , where $P_\mathrm{el}=J(V)\times V$ is the extracted electrical power density, and $P_\mathrm{phot}$ is the photonic heat exchange between the emitter and the cell, given by $P_\mathrm{phot}=P_\mathrm{\omega<\omega_\mathrm{g}}+P_\mathrm{\omega>\omega_\mathrm{g}}$, where the terms $P_\mathrm{\omega<\omega_\mathrm{g}}$ and $P_\mathrm{\omega>\omega_\mathrm{g}}$ correspond, respectively, to heat exchange below-, and above-band gap. These are given by $P_\mathrm{\omega<\omega_\mathrm{g}}=Q_\mathrm{e,\omega<\omega_\mathrm{g}}-Q_\mathrm{o,\omega<\omega_\mathrm{g}}=$ and $P_\mathrm{\omega>\omega_\mathrm{g}}=Q_\mathrm{e,\omega>\omega_\mathrm{g}}-Q_\mathrm{o,\omega>\omega_\mathrm{g}}e^{qV/kT_\mathrm{C}}$, where:
\begin{equation}\label{eq:6}
Q_\mathrm{o/e,\omega<\omega_\mathrm{g}}=\frac{1}{4\pi^2}\int_{0}^{\omega_\mathrm{g}} \hbar\omega \Phi(\omega) n(\omega,T_\mathrm{C/H}) d\omega
\end{equation}
and
\begin{equation}\label{eq:6b}
Q_\mathrm{o/e,\omega>\omega_\mathrm{g}}=\frac{1}{4\pi^2}\int_{\omega_\mathrm{g}}^{\infty} \hbar\omega \Phi(\omega) n(\omega,T_\mathrm{C/H}) d\omega.
\end{equation}
The term $P_\mathrm{\omega<\omega_\mathrm{g}}$ becomes important for PV cells made of polar semiconductors \cite{Chen_FanNFTPV2015}.}

\par{In the absence of an applied bias in the PV cell, the non-radiative generation and recombination currents in Eq. \ref{eq:1} are balanced, hence, the short-circuit current, ($J_\mathrm{sc}=J(V = 0)$) is:
\begin{equation}\label{eq:4}
J_\mathrm{sc}=J_\mathrm{e}-J_\mathrm{o}.
\end{equation}
Following Shockley-Queisser analysis \cite{DetailedBalance1961}, we first use a simple model that facilitates analytic derivation by assuming $R(V)=R_\mathrm{o}e^{qV/kT_\mathrm{C}}$. Then, based on Eq. \ref{eq:1}, the open-circuit voltage ($J(V_\mathrm{oc}) = 0$) becomes: 
\begin{equation}\label{eq:5}
V_\mathrm{oc}=\frac{kT_\mathrm{C}}{q}\mathrm{ln}[\frac{J_\mathrm{e}+R_\mathrm{o}}{J_\mathrm{o}+R_\mathrm{o}}].
\end{equation}
It is generally desirable to increase both $J_\mathrm{sc}$ and $V_\mathrm{oc}$, which increases the power density. Below, we show that increasing $V_\mathrm{oc}$ also increases the efficiency.}

\par{Assuming a normalized emission spectrum that is non-zero only within the frequency range $[\omega_\mathrm{g}, \omega_\mathrm{g}+\delta\omega]$, and ignoring below-band gap heat exchange ($P_\mathrm{\omega<\omega_\mathrm{g}}=0$), the upper limit of integration in Eqs. \ref{eq:2} and \ref{eq:6b} becomes $\omega_\mathrm{g}+\delta\omega$. By further assuming that $\omega_\mathrm{g}\gg kT_\mathrm{H}$, $\delta\omega\ll kT_\mathrm{C}$, and $\Phi(\omega)$ is slowly varying within the range of $[\omega_\mathrm{g}, \omega_\mathrm{g}+\delta\omega]$, to the lowest order of $\delta\omega$, the integrations in Eqs. \ref{eq:2} and \ref{eq:6b} can be evaluated to give:
\begin{equation}\label{eq:7}
J_\mathrm{o/e}=\frac{q}{4\pi^2}\delta\omega\Phi(\omega_\mathrm{g})e^{-\omega_\mathrm{g}/kT_\mathrm{C/H}}
\end{equation}
and
\begin{equation}\label{eq:8}
Q_\mathrm{o/e,\omega>\omega_\mathrm{g}}=J_\mathrm{o/e}\frac{\hbar}{q}(\omega_\mathrm{g}+\delta\omega).
\end{equation}
Then, the efficiency approaches:
\begin{equation}\label{eq:9}
\eta\approx\frac{qV}{\hbar(\omega_\mathrm{g}+\delta\omega)}
\end{equation}
suggesting that the maximum efficiency corresponds to the open-circuit voltage, $V_\mathrm{oc}$.}

\par{We now consider two regimes, the radiative regime, where radiative recombination dominates, i.e. $R_\mathrm{o}\ll J_\mathrm{o}$, and the non-radiative region, where non-radiative recombination dominates, i.e. $R_\mathrm{o}\gg J_\mathrm{o}$. In the radiative regime, from Eqs. \ref{eq:5} and Eq. \ref{eq:7} we obtain $V_\mathrm{oc}=(\hbar \omega_\mathrm{g}/q)(1-T_\mathrm{C}/T_\mathrm{H})$. Correspondingly, for a monochromatic spectrum, i.e. for $\delta\omega\rightarrow 0$, Eq. \ref{eq:9} yields $\eta=1-T_\mathrm{C}/T_\mathrm{H}$ \cite{Harder_2003,BuddhirajuE3609,NAM2014287}. This is the Carnot thermodynamic efficiency limit of a heat engine. When $\delta\omega$ increases from $0$, the efficiency decreases, as can be seen in Eq. \ref{eq:9}.}

\par{In contrast to the radiative regime, however, many semiconducting materials used in TPV systems are subject to substantial non-radiative recombination. For these semiconductors, the non-radiative regime is more relevant. In this case, it is easy to see from Eqs. \ref{eq:4} and \ref{eq:7} that the short-circuit current, $J_\mathrm{sc}$, increases linearly with $\delta\omega$. In the non-radiative limit, the $J_\mathrm{o}$ term in the denominator of Eq. \ref{eq:5} becomes irrelevant and $V_\mathrm{oc}$ can be written as:
\begin{equation}\label{eq:10}
V_\mathrm{oc}=\frac{kT_\mathrm{C}}{q}\mathrm{ln}[\frac{\delta\omega}{R^{'}_\mathrm{o}}+1].
\end{equation}
where $R^{'}_\mathrm{o}=(4\pi^2 R_\mathrm{o}e^{\hbar \omega_\mathrm{g}/kT_\mathrm{H}})/(q \Phi(\omega_\mathrm{g}))$. Therefore, the open-circuit voltage increases with the bandwidth $\delta\omega$. Since both the short-circuit current and open-circuit voltage increase with bandwidth, we expect that the electrical power density also increases with the exchange spectrum bandwidth. Furthermore, the same can be said for the efficiency. Particularly, Eqs. \ref{eq:9} and \ref{eq:10} yield:
\begin{equation}\label{eq:11}
\eta\approx\frac{kT_\mathrm{C}}{\hbar(\omega_\mathrm{g}+\delta\omega)}\mathrm{ln}[\frac{\delta\omega}{R^{'}_\mathrm{o}}+1].
\end{equation}
One can show that $\eta$ is an increasing function of $\delta\omega$ as long as $\delta\omega\ll \omega_\mathrm{g}$. In this limit, therefore, we show that simultaneous increase in efficiency and power density can be achieved by broadening the bandwidth of the normalized emission spectrum. So far, our treatment pertains to both far-field and near-field TPVs. In what follows, we focus on near-field TPVs due to significantly better performance in terms of both power density and efficiency, compared to far-field TPVs \cite{Greffet_NFTPV2006,Abdallah_ReviewNFTPV,Zhao_NFTPV}.}

\begin{figure*}[]
\centering
\includegraphics[width=1\linewidth]{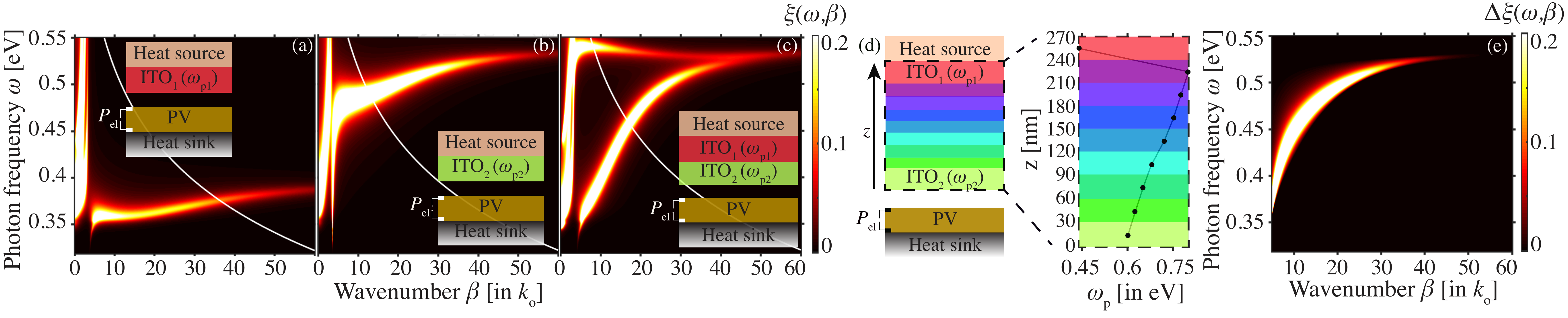}
  \caption{Photon transmission probability $\xi(\omega,\beta)$ for (a) a conventional near-field TPV consisting of a $400$ nm InAs cell and a single-layer $30$ nm ITO emitter with $\omega_\mathrm{p1}=0.44$ eV, at a vacuum gap distance of $d=10$ nm. (b) Same as (a) for $\omega_\mathrm{p2}=0.6$ eV. (c) Bi-layer emitter case with $\omega_\mathrm{p1}=0.44$ eV and $\omega_\mathrm{p2}=0.6$ eV. The white curve in (a)-(c) corresponds to the Planck distribution $n(\omega,T_\mathrm{H}=1300$ K$)$. (d) Left-schematic of a design strategy for further enhancing heat transfer by by inserting a plasmonic heterostructure between the top and bottom emitter layers in (c). The plasma frequency distribution $\omega_\mathrm{p}(z)$ in the heterostructure is shown on the right. (e) Difference between the photon transmission probability for the schematic of panel (d) and that of panel (c). The optical constants for all constituent materials have been taken as in \cite{Zhao_NFTPV}, except for the damping rate  of ITO, which was reduced by an order of magnitude for demonstration purposes. The $x$-axis of the contour plots is normalized to the free-space wavenumber $k_\mathrm{o}=\omega/c$.}
 \label{fig:Figure2} 
\end{figure*}

\par{Motivated by the discussion above, below, we outline a practical design strategy for broadening the normalized emission spectrum $\Phi(\omega)$ in an exemplary near-field TPV system.  The system consists of an InAs cell facing a plasmonic indium tin oxide (ITO) emitter as shown in the inset of Fig. \ref{fig:Figure2}a. We describe the permittivity of ITO with a Drude model $\epsilon(\omega)=\epsilon_\mathrm{\infty}(1-\omega_\mathrm{p}^2/(\omega^2+i\omega\gamma))$, where $\epsilon_\mathrm{\infty}=4$, and $\gamma$ is the damping rate of ITO set to $\gamma=0.1$ eV \cite{Zhao_NFTPV, Papadakis_TunableNFHT, Brongersma_ITO2013}. As a heat source, we consider a tungsten back-side reflector that uniformly spreads thermal energy to the ITO surface area. To ensure recycling of low-energy photons, a perfect electric conductor layer is placed at the back side of the cell (heat sink in the inset of Fig. \ref{fig:Figure2}a) \cite{Yablonovich_PNASTPV2019}. This motif has been previously shown to achieve the largest reported combination of efficiency and power density in simulations \cite{Zhao_NFTPV}. Similar to \cite{Zhao_NFTPV}, we consider a separation distance of $d=10$ nm between emitter and the cell.}

\par{To illustrate the proposed scheme for broadening the normalized emission spectrum bandwidth, in Fig. \ref{fig:Figure2} we reduce the damping rate of ITO by one order of magnitude, and discuss in what follows the photon transmission probability, $\xi(\omega,\beta)$. For the remaining of the paper, the damping rate of ITO is restored to its realistic value of $\gamma=0.1$ eV. As seen in Fig. \ref{fig:Figure2}a, where the plasma frequency of the ITO emitter is set to $\omega_\mathrm{p1}=0.44$ eV, the heat exchange is relatively narrow in bandwidth. Increasing the plasma frequency to $\omega_\mathrm{p2}=0.6$ eV blueshifts the frequency range where heat transfer is maximized, nevertheless the bandwidth remains narrow, as shown in Fig. \ref{fig:Figure2}b. In order to broaden the bandwidth, we consider a bi-layer emitter composed of the two previous layers with plasma frequencies $\omega_\mathrm{p1}$ and $\omega_\mathrm{p2}$, where the layer with the larger plasma frequency ($\omega_\mathrm{p2}$) is placed closer to the vacuum gap, as can be seen in the inset of Fig. \ref{fig:Figure2}c. The bandwidth of the heat exchange in this bi-layer emitter case has considerably increased, which can be understood as follows: at large in-plane wavenumbers, $\beta\rightarrow\infty$, evanescent modes that tunnel from the bi-layer emitter to the cell have little penetration into the top layer ($\omega_\mathrm{p1}$), therefore only the layer closest to the vacuum gap contributes to heat transfer, for which heat transfer becomes prominent at high frequencies approaching $\omega_\mathrm{p2}$. Analogously, at small wavenumbers, $\beta\rightarrow0$, the field expands across the whole depth of the bi-layer emitter, thereby allowing for considerable amount of heat exchange to occur via modes supported by the top layer (red layer in Fig. \ref{fig:Figure2}c). Hence, the point of maximum heat transfer (maximum $\xi(\omega,\beta)$) redshifts towards $\omega_\mathrm{p1}$ as $\beta$ decreases. The heat exchange bandwidth in Fig. \ref{fig:Figure2}c is nearly $[\omega_\mathrm{p1},\omega_\mathrm{p2}]$, and can be engineered on demand through the selection of $\omega_\mathrm{p1}$, $\omega_\mathrm{p2}$.}

\par{Building upon the concept of the bandwidth broadening discussed in Figs. \ref{fig:Figure2}a-\ref{fig:Figure2}c, the amount of heat exchange between the emitter and the cell within this bandwidth can be further increased. To achieve this, we insert a heterostructure composed of thin plasmonic layers in between the top ($\omega_\mathrm{p1}$) and bottom ($\omega_\mathrm{p2}$) layers, as shown in the schematic of Fig. \ref{fig:Figure2}d, where the plasma frequency profile in this heterostructure, $\omega_\mathrm{p}(z)$, is shown on the right. With this modification, each plasmonic film with $\omega_\mathrm{p,i}$, for $i=1,2,..$, provides additional contributions to heat transfer due to supported plasmonic modes occurring at energies lower than $\omega_\mathrm{p,i}$ \cite{Rodriguez_Optimization2017}. Since the absolute wavenumber $\beta$ of a plasmonic mode scales with its respective plasma frequency $\omega_\mathrm{p,i}$, decreasing the plasma frequency as one approaches the vacuum gap maximizes the contribution of each layer to heat transfer. In Fig. \ref{fig:Figure2}e we show the difference between the photon transmission probability of the optimized structure (schematic in Fig. \ref{fig:Figure2}d) from the bi-layer emitter of Fig. \ref{fig:Figure2}c. The bright regions in the $(\omega,\beta)$ plane show the additional contributions to heat transfer, enabled by the inserted plasmonic heterostructure. The plasma frequency profile shown in Fig. \ref{fig:Figure2}d can be obtained, for example, through a gradual change in the doping profile in the ITO region, which can be achieved by gradually altering the conditions (e.g. oxygen concentration and temperature \cite{Lee_Plasmonstor, Jeong_2008}, or pressure \cite{Chen2016EffectOS}) during thin film deposition.}

\par{Assuming that one has the flexibilty to select the band gap of the semiconductor for the PV cell, the selection of the bi-layer emitter's plasma frequencies, $\omega_\mathrm{p1}$ and $\omega_\mathrm{p2}$, in Fig. \ref{fig:Figure2}c, depends critically on the temperature of the emitter, $T_\mathrm{H}$. The normalized emission spectra for the structures in Figs. \ref{fig:Figure2}c and \ref{fig:Figure2}e lie largely in the interval $[\omega_\mathrm{p1}, \omega_\mathrm{p2}]$. The Planck distribution $n(\omega, T_\mathrm{H})$ at the emitter temperature $T_\mathrm{H}$ decreases with increasing $\omega$ and becomes negligible when $\omega \gg k T_\mathrm{H}$, whereas the density of the states associated with heat exchange increases with frequency in a slower rate. Combining these two considerations, one should, therefore, select the plasma frequencies to be slighlty larger than $k T_\mathrm{H}/\hbar$. Our choice of the plasma frequency $\omega_\mathrm{p1}=0.44$ eV in Figs. \ref{fig:Figure2}a, \ref{fig:Figure2}c is for an emitter with a temperature $T_\mathrm{H}=1300$ K, as can be seen by the white curve which corresponds to $n(\omega,T_\mathrm{H}=1300$ K$)$. The plasma frequency $\omega_\mathrm{p2}=0.6$ eV in Figs. \ref{fig:Figure2}b, \ref{fig:Figure2}c was selected in order to clearly visualize the bandwidth broadening mechanism in the low-loss limit. However, for emitter temperatures in the range of $T_\mathrm{H}=1300$ K, in what follows we reduce  $\omega_\mathrm{p2}$ to $0.5$ eV.}

\par{We now show that the bandwidth broadening strategy as shown in Fig. \ref{fig:Figure2} can indeed be used to improve the performance of near-field TPV systems. As a benchmark, we consider the single-layer emitter near-field TPV displayed in the schematic of Fig. \ref{fig:Figure2}a, for an InAs cell, and an ITO plasmonic emitter with $\omega_\mathrm{p1}=0.44$ eV and $\gamma=0.1$ eV. In the following results, we account for non-idealities in the materials, namely, below-band gap absorption (as outlined in the formalism above, see Eq. \ref{eq:6}) and non-radiative recombination in InAs.}

\par{Regarding non-radiative processes in InAs, the Shockley-Read-Hall (SRH) and Auger non-radiative mechanisms are dominant, which constitute the first and second terms, respectively, in \cite{Kfaifeng_Auger}:
\begin{equation}\label{eq:12}
R(V)=\frac{pn-n_\mathrm{i}^2}{\tau(n+p+2n_\mathrm{i})}t_\mathrm{c}+(C_\mathrm{p}p+C_\mathrm{n}n)(np-n_\mathrm{i}^2)t_\mathrm{c}.
\end{equation}
In Eq. \ref{eq:12}, $t_\mathrm{c}$ is the thickness of the cell, $\tau$ is the SRH lifetime, $C_\mathrm{n}$ and $C_\mathrm{p}$ are the Auger coefficients, $n_\mathrm{i}$ is the intrinsic carrier concentration in the cell, and $n$ and $p$ are the electron and hole densities, respectively. We set the thickness of the InAs cell to $t_\mathrm{c}=400$ nm (similar to Fig. \ref{fig:Figure2}), which maximizes power density in the presence of non-radiative recombination \cite{Zhao_NFTPV}.}

\par
\begin{figure}[]
\centering
\includegraphics[width=\linewidth]{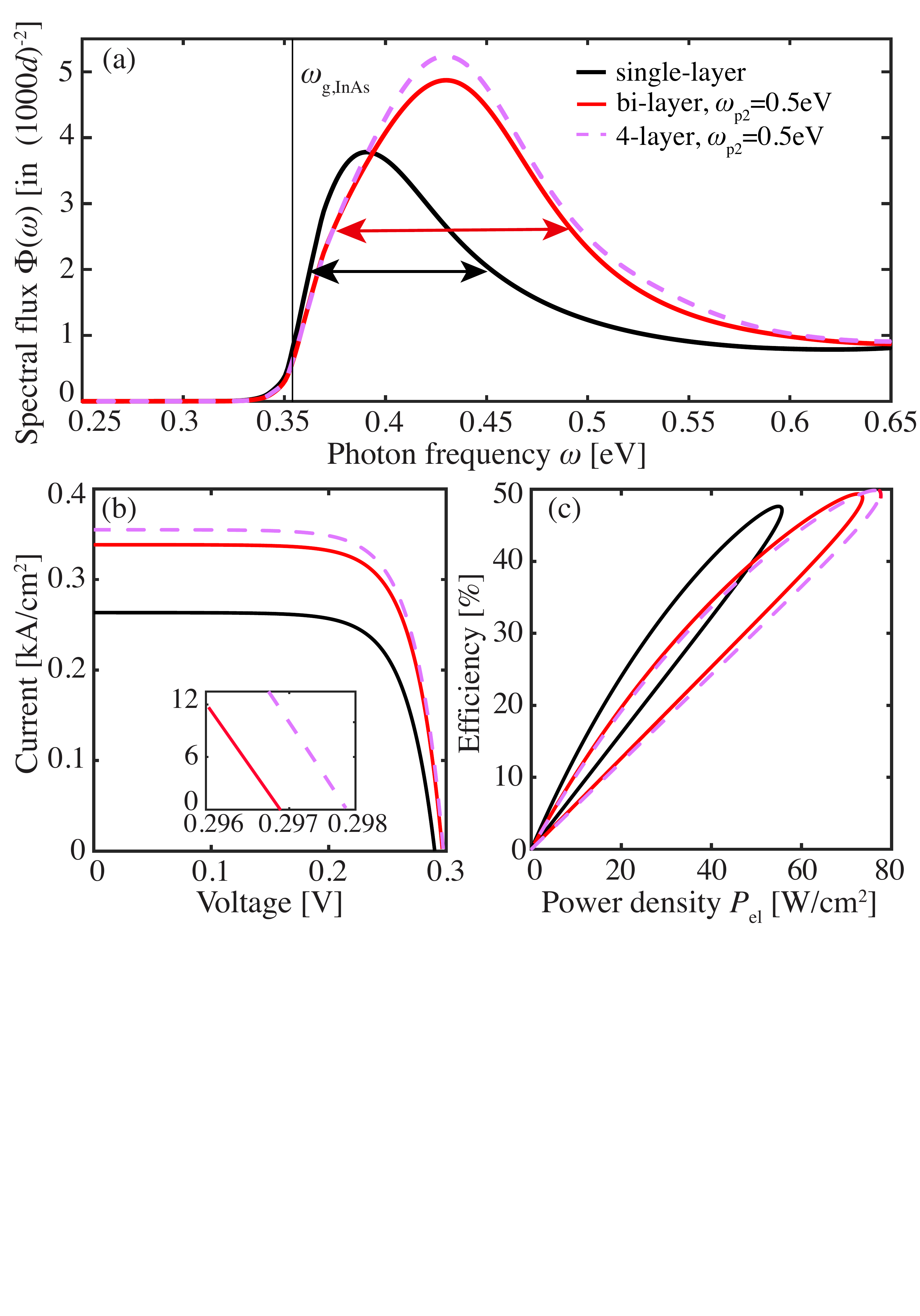}
  \caption{(a) Normalized emission spectrum, $\Phi(\omega)$, for a near-field TPV with an InAs cell and an ITO emitter. The single-layer emitter case pertains to the schematic in the inset of Fig. \ref{eq:1}a, with $\omega_\mathrm{p1}=0.44$ eV. The bi-layer case pertains to the schematic in the inset of Fig. \ref{fig:Figure2}c for $\omega_\mathrm{p2}=0.5$ eV. The arrows indicate the bandwidth of the corresponding normalized emission spectra. The four-layer emitter case corresponds to implementing the design strategy of Fig. \ref{fig:Figure2}d, with the intermediate layers having $\omega_\mathrm{p3}=0.6$ eV and $\omega_\mathrm{p4}=0.575$ eV. (b) Current-voltage characteristics. The inset magnifies the range where current is near-zero. The voltage at zero current is the open-circuit voltage, $V_\mathrm{oc}$. (c) efficiency $\eta$ vs power density $P_\mathrm{el}$ showing simultaneous increase in $\eta$ and $P_\mathrm{el}$ as the number of layers increases.}
 \label{fig:Figure3}
\end{figure}

{We set the cell temperature at $T_\mathrm{C}=300$ K and the emitter temperature at $T_\mathrm{H}=1300$ K. In Fig. \ref{fig:Figure3}a, we display the normalized emission spectrum $\Phi(\omega)$ for the single-layer emitter configuration with the black curve, and its bandwidth (full width at half maximum) is shown with the black arrow. By considering the bi-layer emitter, as shown in the schematic of Fig. \ref{fig:Figure2}c, with $\omega_\mathrm{p2}=0.5$ eV, the spectrum broadens considerably, as can be seen with the red curve and arrow in Fig. \ref{fig:Figure3}a. In Fig. \ref{fig:Figure3}b we show that this bandwidth broadening indeed yields simultaneous increase in short-circuit current and open-circuit voltage.}

\par{Next, we implement the strategy outlined in Figs. \ref{fig:Figure2}d, e for further increasing the heat transfer, by inserting intermediate plasmonic layers in between the layers with $\omega_\mathrm{p1}$ and $\omega_\mathrm{p2}$. Here, we consider two intermediate layers with $\omega_\mathrm{p3}=0.6$ eV and $\omega_\mathrm{p4}=0.575$ eV in the sequence displayed in Fig. \ref{fig:Figure2}d. With these additional layers, the bandwidth of the normalized emission spectrum does not change significantly compared to the two-layer structure, however, its amplitude increases. In Fig. \ref{fig:Figure3}b we confirm that adding these intermediate layers in the emitter structure yields indeed further increase in short-circuit current and open-circuit voltage.}

\par{In Fig. \ref{fig:Figure3}(c) we display the efficiency, $\eta$, versus power density, $P_\mathrm{el}$, obtained by tuning the load voltage $V$ from 0 to $V_\mathrm{oc}$. Increasing the number of emitter layers with respect to the single-layer emitter (black curve) yields simultaneous increase in power density and efficiency, in consistency with the increase in $I_\mathrm{sc}$ and $V_\mathrm{oc}$, respectively, as discussed above.}

\begin{figure}[h!]
\centering
\includegraphics[width=\linewidth]{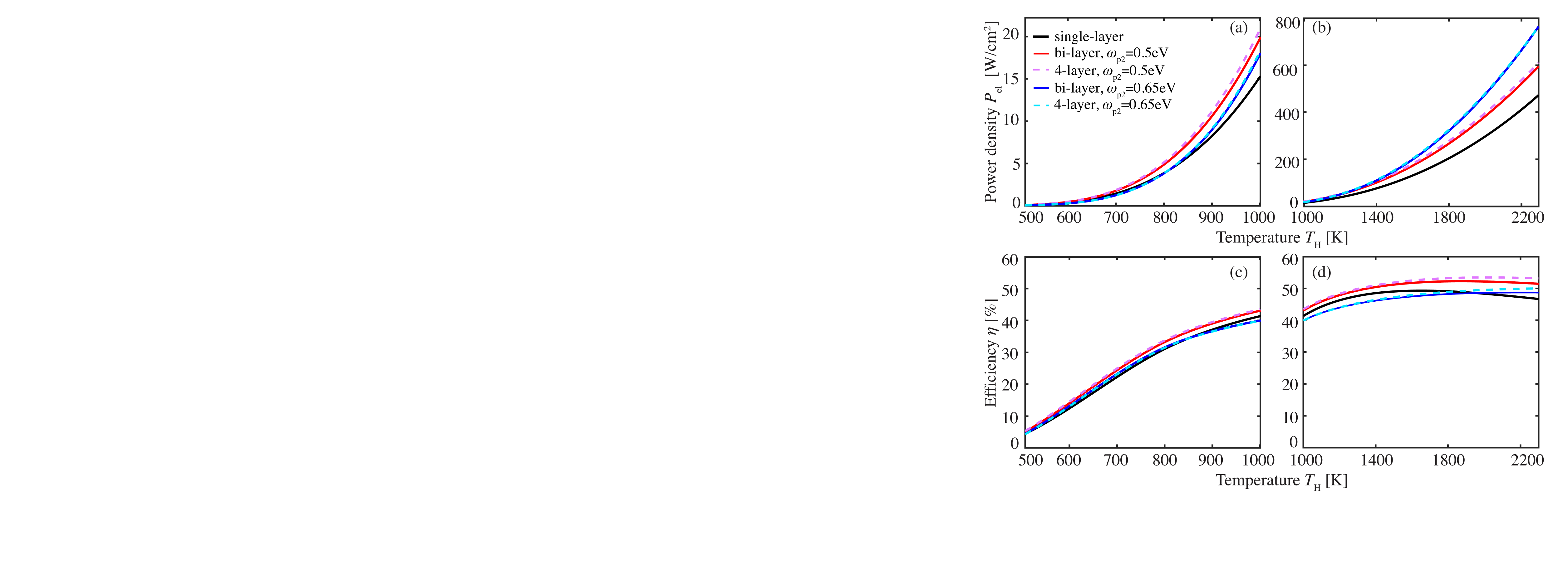}
  \caption{(a), (b) Power density $P_\mathrm{el}$ and (c), (b) efficiency $\eta$ for the single-layer emitter case, and for the bi-layer emitter and four-layer cases as a function of $T_\mathrm{H}$. Panels (a), (c) correspond to the waste heat temperature range, whereas panels (b), (d) correspond to higher emitter temperatures. The solid red and dashed pink curves correspond to a selection of $\omega_\mathrm{p2}=0.5$ eV, for optimizing performance near the waste heat temperature range, same as in Fig. \ref{fig:Figure3}. The solid blue and dashed cyan curves correspond to $\omega_\mathrm{p2}=0.65$ eV for optimizing performance at higher temperatures.}
 \label{fig:Figure4}
\end{figure}

\par{In Fig. \ref{fig:Figure4}, we study the performance of the near-field TPVs discussed in Figs. \ref{fig:Figure3} for a wide range of emitter temperatures from $T_\mathrm{H}=500$ K to the melting point of ITO at $T_\mathrm{H}=2200$ K. Particularly, we show the power density (Figs. \ref{fig:Figure4}a, b) and efficiency (Figs. \ref{fig:Figure4}c, d) for the single-layer emitter configuration (black curves) and compare with the bi-layer emitter case (red curves) for $\omega_\mathrm{p2}=0.5$ eV, considered in Fig. \ref{fig:Figure3}, and the corresponding four-layer emitter case (dashed pink curves). In terms of power density, the bi-layer yields significant improvement with respect to the single-layer emitter, in the temperature range $T_\mathrm{H}<1000$ K (Fig. \ref{fig:Figure4}a), which is important for practical applications of TPVs since it corresponds to the waste heat, low-grade temperature range where the largest portion of the energy consumed by the manufacturing sector is rejected into the environment in the form of waste heat \cite{DOE2008, Zhao_NFTPV}. We also present results where we optimize the TPV power density at higher emitter temperatures, shown in Fig. \ref{fig:Figure4}b. In this higher temperature operation, it is important to considerably blueshift the normalized emission spectra $\Phi(\omega)$ shown in Fig. \ref{fig:Figure3}, by increasing $\omega_\mathrm{p2}$ to $\omega_\mathrm{p2}=0.65$ eV. Therefore, for high temperatures, using a bi-layer emitter with $\omega_\mathrm{p2}=0.65$ eV (blue curves), we achieve a significantly higher power density as compared to the case with $\omega_\mathrm{p2}=0.5$ eV, as shown in Fig. \ref{fig:Figure4}b. Further small improvement is possible with the use of a four-layer emitter with  $\omega_\mathrm{p3}=0.75$ eV, and $\omega_\mathrm{p4}=0.725$ eV (dashed cyan curves), due to thermal emission enhancement as discussed in Figs. \ref{fig:Figure2}e, d.}

\par{From Figs. \ref{fig:Figure4}a, b it can be seen that the power density increases with the number of emitter layers for both considered values of $\omega_\mathrm{p2}$. At the lower temperature of $1300$ K, we achieve a maximum power density of $78.8$ W$/\mathrm{cm}^2$ with the four-layer emitter, which corresponds to a power density improvement of $37\%$ with respect to the single-emitter case at the same temperature. Similarly, at the higher temperature of $T_\mathrm{H}=2200$ K, the power density reaches $666$ W$/\mathrm{cm}^2$, corresponding to nearly $62\%$ improvement with respect to the single-layer emitter at the same temperature. For all considered emitters in Figs. \ref{fig:Figure4}a, b, the power density increases as a function of emitter temperature, as expected for TPV systems.}

\par{In terms of efficiency, the reported values of maximum efficiency in Figs. \ref{fig:Figure4}c, d for each case were determined by sweeping the load voltage from $0$ to $V_\mathrm{oc}$. Due to the presence of below-band gap absorption in InAs, the load voltage at which maximum efficiency is achieved is slightly smaller than $V_\mathrm{oc}$ (see Eq. \ref{eq:9}). From Fig. \ref{fig:Figure4}d, we see that the four-layer emitter with $\omega_\mathrm{p2}=0.5$ eV yields a conversion efficiency of $50\%$ at $1300$ K, corresponding to a $5\%$ improvement with respect to the single-layer case, due to a broader normalized emission bandwidth. Similarly, at $T_\mathrm{H}=2200$ K, the efficiency with the four-layer emitter with $\omega_\mathrm{p2}=0.5$ eV reaches $53.3\%$, providing a nearly $13\%$ improvement with respect to the single-layer emitter. We note that for the temperature range $T_\mathrm{H}<1800$ K, the bi-layer and four-layer cases with $\omega_\mathrm{p2}=0.65$ eV (blue and dashed cyan curves) are sub-optimal in terms of efficiency, yielding efficiencies smaller than that with the single-layer emitter, as shown in Fig. \ref{fig:Figure4}c. In this range of temperatures with a larger $\omega_\mathrm{p2}$, the bandwidth broadening provided by the additional emitter layers is not fully exploitable due to the misalignement between the range where there is significant photon number in the Planck distribution and $\Phi(\omega)$. Particularly, the Planck distribution is redshifted with regards to the plasmonic resonance of the emitter near $\omega_\mathrm{p2}$. The selection of $\omega_\mathrm{p2}$ should be made according to the targeted emitter temperature.}

\par{As can be seen in Figs. \ref{fig:Figure4}c, d, the efficiency increases rapidly as a function of temperature for small emitter temperatures, while saturating at approximately $45-50$ \% for higher temperatures, and reduces at very high temperatures outside the range displayed in Fig. \ref{fig:Figure4}d \cite{Harder_2003}. This can be understood as follows. For $kT_\mathrm{H}<\omega_\mathrm{g}$, the frequency range in which the Planck distribution contains a photon number that is significantly above zero lies at energies smaller than the peak of the normalized emission spectrum, $\Phi(\omega)$. Hence, the amount of thermal photon flux received by the cell (see Eq. \ref{eq:6b}) is not optimized to energies around its band gap, therefore the cell under-performs. As $T_\mathrm{H}$ increases, the alignment between $n(\omega,T)$ and $\Phi(\omega)$ improves, and less photonic flux is required for the cell to generate electrical power, and thereby the efficiency increases, up to the point at which the frequency range where $n(\omega,T)$ is significantly above zero extends to energies much larger than the band gap, for which its misalignment with $\Phi(\omega)$ leads to efficiency decrease, at $T_\mathrm{H}\gg 2200$ K.}

\begin{figure}[h!]
\centering
\includegraphics[width=\linewidth]{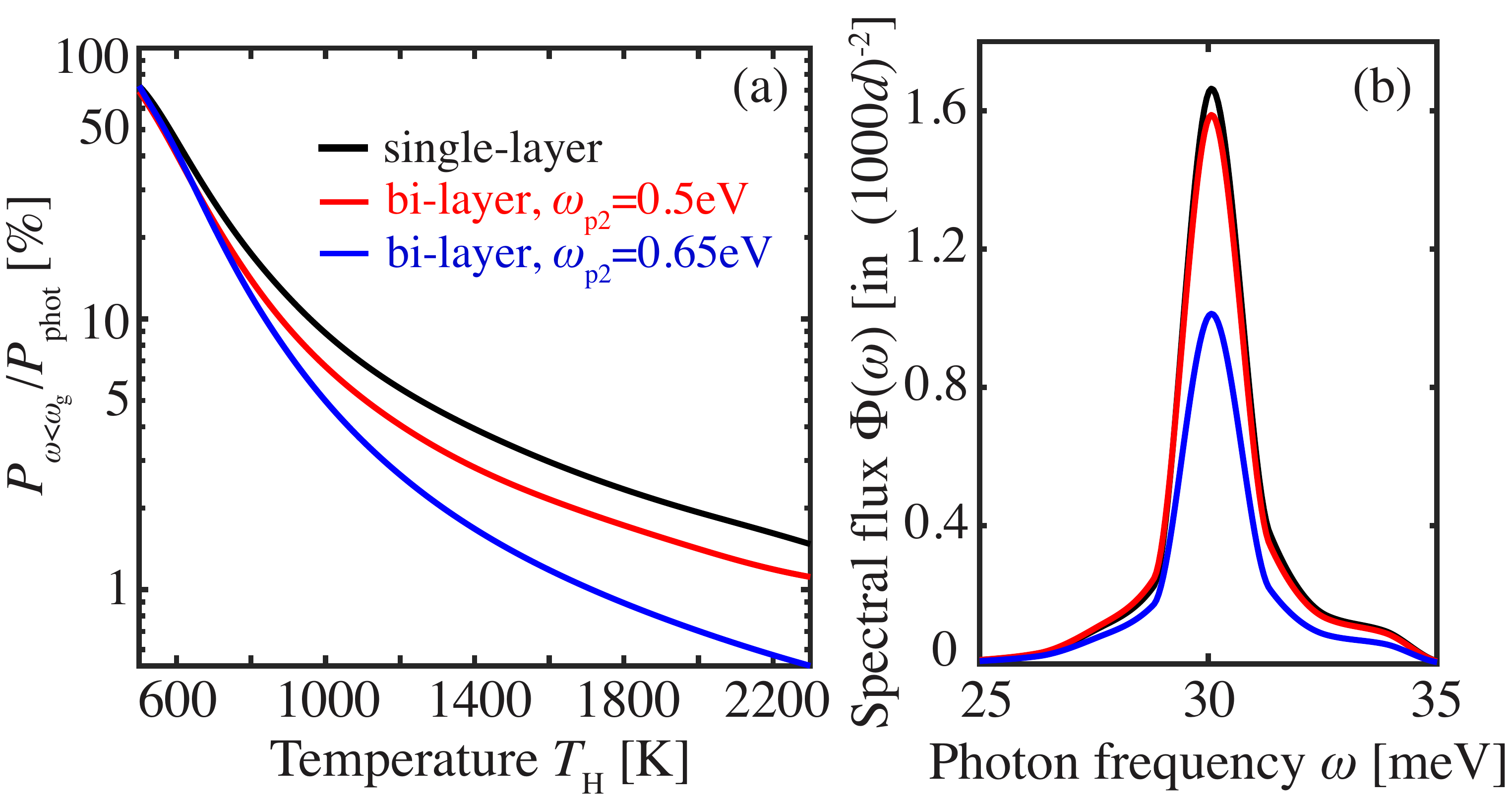}
  \caption{(a) Fraction of below-band gap heat exchange for the same results as in Fig. \ref{fig:Figure4}, for an applied bias of $V=V_\mathrm{oc}$. (b) Below-band gap normalized emission spectrum, $\Phi(\omega)$, near the Reststrahlen band of InAs. Black curve corresponds to single-layer emitter case, while blue and red ones correspond to the bi-layer emitter with $\omega_\mathrm{p2}=0.5$ eV and $0.65$ eV, respectively.}
 \label{fig:Figure5}
\end{figure}

\par{In the TPV system as considered above, the below-band gap absorption, which is parasitic, plays an important role in determining performance. Fundamentally, below-band gap heat exchange can be completely suppressed with the use of non-polar semiconductors \cite{Chen_FanNFTPV2015}. Nevertheless, this restricts the range of available band gaps and suitable emitter temperatures for TPV systems. In contrast, all polar materials exhibit below-band gap absorption concentrated near their Reststrahlen band \cite{Cadwell_SiC,Papadakis_polaritons}. In this band, surface phonon polariton modes parasitically contribute to near-field heat transfer, leading to efficiency reduction in near-field TPVs \cite{Chen_FanNFTPV2015}. For InAs, the Reststrahlen band lies at $30$ meV. The resulting below-band gap absorption can have significant impact on the performance of the near-field TPV systems that we consider. For example, by calculating the ratio of below-band gap heat exchange, $P_\mathrm{\omega<\omega_\mathrm{g}}$, to the total photonic power exchange, $P_\mathrm{phot}$, at $V=V_\mathrm{oc}$ (see Eq. \ref{eq:6}), we show in Fig. \ref{fig:Figure5}a that, for the single-layer emitter configuration, below-band gap absorption constitutes nearly $60\%$ of the total heat flux at low emitter temperatures, as shown for $T_\mathrm{H}=500$ K.}

\par{Nevertheless, we show in Fig. \ref{fig:Figure5} that below-band gap heat exchange can be considerably reduced at all emitter temperatures, even for polar semiconductors like InAs, with the proposed bi-layer TPV emitters considered in Fig. \ref{fig:Figure4}. In Fig. \ref{fig:Figure5}b, we compute the below-band gap normalized emission spectrum, $\Phi(\omega)$, near the Reststrahlen band of InAs. It can be seen that below-band gap photon flux is reduced in amplitude for the bi-layer emitters compared to the single-layer emitter. The flux further reduces as $\omega_\mathrm{p2}$ increases from $\omega_\mathrm{p2}=0.5$ eV (red curve) to $\omega_\mathrm{p2}=0.65$ eV (blue curve). In this range of frequencies, heat transfer is dominated by the coupling between the surface phonon polariton mode of the InAs cell and the plasmonic mode of the ITO layer closest to the vacuum gap. At a given wavevector $\beta$, the strongest coupling occurs when the two modes have similar frequencies. Consequently, we expect maximum coupling for equal dielectric permittivity between the cell and the ITO layer closest to the vacuum gap. The permittivity of InAs at $30$ meV is nearly $-18$. For $\omega_\mathrm{p1}=0.44$ eV (black curve), the real part of the dielectric permittivity of ITO at $30$ meV is approximately $-30$, whereas for the bi-layer emitter with $\omega_\mathrm{p2}=0.5$ eV (red curve) it is approximately $-90$, and reduces further to $-150$ for $\omega_\mathrm{p2}=0.65$ eV (blue curve). Therefore, the coupling between the emitter and the cell weakens with the use of a bi-layer emitter, and as we increase $\omega_\mathrm{p2}$. For example, for $\omega_\mathrm{p2}=0.65$ eV, it can be seen in Fig. \ref{fig:Figure5}a that below-band gap absorption can be as small as $2\%$ and $0.57\%$ of the total power exchange, at $T_\mathrm{H}=1300$ K and $T_\mathrm{H}=2200$ K, respectively. The overall reduction of below-band gap heat flux in Fig. \ref{fig:Figure5} contributes to the large efficiency values for the bi-layer emitters shown in Fig. \ref{fig:Figure4}.}
\par{The fraction of below-band gap absorption decreases with emitter temperature, as shown in Fig. \ref{fig:Figure5}a for all emitter cases. This can be understood since the spectrum of the Planck distribution, $n(\omega,T_\mathrm{H})$, blueshifts with temperature. In contrast, the below-band gap peak of the normalized emission, $\Phi(\omega)$, shown in Fig. \ref{fig:Figure5}b, is fixed at the Reststrahlen band of InAs. Hence, the spectral overlap between these two functions decreases with increasing emitter temperature, yielding a reduced integrated photonic heat flux below-band gap, $Q_\mathrm{e,\omega<\omega_\mathrm{g}}$ (Eq. \ref{eq:6}), thereby reducing the overall heat transfer at frequencies below-band gap.}

\par{In conclusion, we have shown that broadening the normalized emission spectrum in TPV systems can lead to simultaneous increase in the short-circuit current and open-circuit voltage in the presence of considerable non-radiative recombination. In turn, this leads to a simultaneous improvement in power density and efficiency. We further outlined a nanophotonic design strategy for practically broadening the near-field emission spectrum, by using a bi-layer plasmonic emitter with an appropriately selected combination of plasma frequencies. The performance of near-field TPVs can be further increased by creating a gradient doping profile, readily achievable with thin film deposition of a single material that is carrier density-tunable. Using a realistic set of material parameters and considering non-idealities (thermalization, below-band gap absorption and non-radiative recombination), we showed significant enhancement in the power density of near-field TPVs, with respect to the conventional single-layer emitter, in addition to efficiency increase and suppression of parasitic below-band gap absorption.}\\

\noindent \textbf{ORCID}\\
Georgia T. Papadakis: 0000-0001-8107-9221\\
%Bo Zhao: 0000-0002-3648-6183\\
Shanhui Fan: 0000-0002-0081-9732\\
 
\noindent \textbf{Notes}\\
The computational package used for near-field heat transfer calculations can be found in \cite{CHEN2018163}. The authors declare no competing financial interest.  We acknowledge the support from the Department of Energy ``Photonics at Thermodynamic Limits'' Energy Frontier Research Center under Grant No. DE-SC0019140. G.T. P. acknowledges the TomKat Postdoctoral Fellowship in Sustainable Energy at Stanford University.\\
 
\providecommand{\noopsort}[1]{}\providecommand{\singleletter}[1]{#1}%

\end{document}